\documentclass[aps,onecolumn,pra,showpacs,amsmath,amssymb,showkeys,floatfix,10pt]{revtex4-2}
\usepackage{graphicx}
\usepackage{epstopdf}
\usepackage{diagbox}
\usepackage{xcolor}
\usepackage{ulem}
\pdfoutput=1
\usepackage{braket}
\usepackage{hyperref}
\usepackage{longtable}
\usepackage{subfigure}
\usepackage{placeins}
\usepackage{float}


\begin{document}

	\title{Zeno and Anti-Zeno Effects in Dark-State Dynamics Under Thermal Dephasing: A Numerical Study}

	\author{Ran Chen}
	\thanks{These authors contributed equally to this work.}
	\affiliation{Faculty of Computational Mathematics and Cybernetics, Lomonosov Moscow State University, Vorobyovy Gory 1, Moscow, 119991, Russia}
  
	\author{Jiangchuan You}
	\thanks{These authors contributed equally to this work.}
	\affiliation{Faculty of Computational Mathematics and Cybernetics, Lomonosov Moscow State University, Vorobyovy Gory 1, Moscow, 119991, Russia}
	
	\author{Alexey Vladimirovich Kulagin}
	\thanks{These authors contributed equally to this work.}
	\affiliation{Faculty of Computational Mathematics and Cybernetics, Lomonosov Moscow State University, Vorobyovy Gory 1, Moscow, 119991, Russia}
	
	\author{Hui-hui Miao}
	\email[Correspondence to: Vorobyovy Gory 1, Moscow, 119991, Russia. Email address: ]{hhmiao@cs.msu.ru (H.-H. Miao)}
	\affiliation{Faculty of Computational Mathematics and Cybernetics, Lomonosov Moscow State University, Vorobyovy Gory 1, Moscow, 119991, Russia}
  
	\author{Yuri Igorevich Ozhigov}
	\email[Correspondence to: Vorobyovy Gory 1, Moscow, 119991, Russia. Email address: ]{ozhigov@cs.msu.ru (Y.I. Ozhigov)}
	\affiliation{Faculty of Computational Mathematics and Cybernetics, Lomonosov Moscow State University, Vorobyovy Gory 1, Moscow, 119991, Russia}
	
	\date{\today}

	\begin{abstract}
	The quantum Zeno and anti-Zeno effects describe how frequent measurements can either suppress or accelerate quantum dynamics. While extensively studied in various platforms, their manifestation in dark-state dynamics remains largely unexplored. Here we investigate the stability of dark states in a cavity quantum electrodynamics (QED) system consisting of two atoms coupled to a single-mode cavity, subject to thermal dephasing that models continuous quantum non-demolition monitoring. Using the Tavis--Cummings model within a Lindblad master equation framework, we perform numerical simulations to investigate how measurement-induced dephasing affects dark-state retention and stabilization time. Through systematic numerical scans, we identify distinct parameter regimes corresponding to Zeno and anti-Zeno behavior: at low dephasing intensities, increasing the measurement strength accelerates the loss of dark-state coherence (anti-Zeno regime), while at higher intensities, it slows down the dynamics and partially recovers dark-state weight (Zeno regime). The transition between these regimes is controlled by the dephasing rates, the cavity photon exchange, and the asymmetry in atom--field couplings. We show that even under strong dephasing, a finite dark-state component persists, demonstrating remarkable robustness. Our results provide insights into the interplay between measurement back-action and decoherence in open quantum systems, with implications for quantum control and information storage.
	\end{abstract}


	\maketitle

	\section{Introduction}
  	\label{sec:Intro}
  	
	Quantum measurement is central to extracting information from microscopic systems. The associated back-action is often accompanied by decoherence. In particular, frequent measurements can suppress transitions between quantum states and freeze the dynamics. This is known as the quantum Zeno effect, first formulated by Misra and Sudarshan~\cite{Misra1977}. Conversely, an inappropriate measurement protocol can accelerate decay, leading to the anti-Zeno regime~\cite{Kaulakys1997}. Both effects have been experimentally demonstrated in various platforms. These include trapped ions~\cite{Itano1990}, cold atoms~\cite{Fischer2001}, nuclear magnetic resonance~\cite{Zheng2013}, optical systems~\cite{Franson2004}, microwave resonators~\cite{Bernu2008}, Bose--Einstein condensates~\cite{Streed2006}, large-spin systems~\cite{Signoles2014}, and superconducting qubits~\cite{Kakuyanagi2015, Harrington2017}. A comprehensive review of Zeno dynamics in open quantum systems is provided by Becker et al.~\cite{Becker2021}. Recent theoretical and experimental studies have further advanced the understanding of Zeno and anti-Zeno effects in various quantum systems~\cite{Chen2021, Li2021, Li2023}.
	
	Previous studies have investigated Zeno and anti-Zeno effects induced by pure dephasing in simpler systems. For instance, Chaudhry and Gong~\cite{Chaudhry2014} analyzed these effects in single two-level systems, while Auffèves et al.~\cite{Auffeves2009, Auffeves2010} studied control by pure dephasing within the Jaynes--Cummings (JC) model. In contrast, our work extends this line of research to a two-atom Tavis--Cummings (TC) model, where the dark-state manifold emerges from the collective coupling of two atoms to a single cavity mode. (In this work, a dark state is defined as a superposition of atomic states that is decoupled from the cavity mode, a notion standard in cavity quantum electrodynamics (QED) and quantum information science.) This allows us to reveal how measurement-induced dephasing affects a decoherence-free subspace, complementing previous single-emitter studies.
  
	Cavity QED provides a natural framework for studying light--matter interaction. The JC model~\cite{Jaynes1963} and the TC model~\cite{Tavis1968} serve as standard starting points. Recent work within these frameworks covers quantum gates~\cite{OzhigovYI2020, Dull2021}, many-body effects~\cite{Smith2021}, entropy~\cite{MiaoHuihui2024}, quantum discord~\cite{MiaoLi2025}, phase transitions~\cite{Prasad2018, Wei2021}, and related topics~\cite{Guo2019, Victorova2020, Kulagin2022, Afanasyev2022, Pluzhnikov2022, Chen2022, LiMiao2024, MiaoOzhigov2024}.
  
  	A dark state is a special quantum superposition. It is decoupled from selected radiative channels and therefore does not absorb or emit photons in the corresponding interaction. Dark states are relevant for various systems and applications. These include sublattice systems~\cite{Chung2024}, superconducting platforms~\cite{Zanner2022}, and energy-storage applications~\cite{Quach2020, Sundar2024}, among many others~\cite{Lee1999, Andre2002, Poltl2012, Tanamoto2012, Hansom2014, Kozyrev2018, Ozhigov2020}. The dark state can be prepared using several techniques. These include electromagnetically induced transparency (EIT)~\cite{Morigi2000, Lu2015, Luo2009}, coherent population trapping (CPT), velocity-selective coherent population trapping (VSCPT), and related methods. In the language of measurement theory, Positive Operator-Valued Measures (POVMs) provide the most general probabilistic description of measurement outcomes. In cavity-QED experiments, continuous monitoring is typically implemented in a quantum non-demolition (QND) manner. Examples include atomic current~\cite{Laflamme2017} and dispersive readout~\cite{Bianchetti2009}. Repeated QND probing can induce Zeno dynamics~\cite{Raimond2010, Raimond2012}. In this paper, we model the measurement back-action as thermal dephasing. This is a dissipative channel that suppresses coherence without directly introducing a new population channel. We numerically study how this dephasing modifies dark-state dynamics in a single-mode cavity. We also identify parameter domains associated with Zeno and anti-Zeno behavior.
  
  	The structure of this article is as follows. Section~\ref{sec:Mulmod} introduces the two-level cavity model and the corresponding Lindblad operators. Section~\ref{sec:Method} describes the numerical integration schemes, specifically the Euler method and the Runge--Kutta method. Section~\ref{sec:Results} presents simulations for two-level systems, including a reduced-basis fidelity analysis. Section~\ref{sec:Conclusion} summarizes the main findings and discusses open problems.
  
\section{Thermal Dephasing of a Pair of Two-Level Atoms in a Single-Mode~Cavity}
  	\label{sec:Mulmod}
  
	Our system consists of a pair of two-level atoms in a single-mode cavity. It also includes photon leakage and replenishment channels. To describe dark-state dynamics, we use the TC model within the rotating-wave approximation (RWA). In ideal symmetric conditions, dark states are decoupled from the cavity mode. To break this decoupling in a controlled way, we include dissipative channels through a Lindblad description~\cite{Davidsson2023}:
  	\begin{equation}
  		\label{eq:QME}
    	   i\hbar\dot{\rho}=[\hat{H},\hat{\rho}]+i\hat{\mathcal{L}}(\hat{\rho})=[\hat{H},\hat{\rho}]+i\sum_{k}\gamma_{k}\left(\hat{A}_{k}\hat{\rho} \hat{A}_{k}^\dagger-\frac{1}{2}\left(\hat{A}_{k}^\dagger\hat{A}_{k}\hat{\rho}+\hat{\rho}\hat{A}_{k}^\dagger\hat{A}_{k}\right)\right).
  	\end{equation}
  	Here, $\hat{H}$ is the Hamiltonian, $\hat{\rho}$ is the density matrix, $\hbar$ is the reduced Planck constant, $\hat{A}_k$ are the Lindblad operators, and $\gamma_k$ are their corresponding rates. Environment-induced dephasing is a standard mechanism in open quantum systems~\cite{Breuer2007}. Depending on the measurement strength, it can lead to Zeno or anti-Zeno behavior~\cite{Ulrich2008}. In our model, $\hat{\mathcal{L}}(\hat{\rho})$ includes both photon exchange with the environment and measurement-induced dephasing channels.

	\subsection{Hamiltonian of a Pair of Two-Level Atoms in a Single-Mode Cavity}
  	\label{sec:Hamiltonian}
  	
  	In this work, we assume exact resonance to isolate the measurement-induced Zeno/anti-Zeno effects from detuning-induced dynamics. Nonzero detuning would break the dark-state decoupling even in the absence of measurement; a systematic study of detuning effects is beyond the present scope and is left for future work.

	The Hamiltonian consists of the following parts:
  	\begin{equation}
  		\label{eq:H_TCM}
    		\hat{H}=\hat{H}_{ph}+\hat{H}_{at}+\hat{H}_{int},
  	\end{equation}
  	where $\hat{H}_{ph}$ corresponds to the photon field, $\hat{H}_{at}$ describes a pair of two-level atoms, and $\hat{H}_{int}$ represents the interaction between the photons and the atoms. The photon Hamiltonian $\hat{H}_{ph}$ is given by:
  	\begin{equation}
  		\label{eq:H_cv}
    		\hat{H}_{ph}=\hbar\omega_{ph}\hat{a}^{\dagger}\hat{a},
  	\end{equation} 
	where $\hbar$ is the reduced Planck constant, $\hat{a}$ and $\hat{a}^{\dagger}$ are the photon annihilation and creation operators, respectively, and $\omega_{ph}$ is the cavity mode frequency. For a pair of two-level atoms, the atomic Hamiltonian is:
	\begin{equation}
		\label{eq:H_at}
    		\hat{H}_{at}=\hbar\omega_{at}\hat{\sigma}_1 ^{\dagger}\hat{\sigma}_1 + \hbar\omega_{at}\hat{\sigma}_2 ^{\dagger}\hat{\sigma}_2.
  	\end{equation} 
	We assume that $|0\rangle_{at}$ and $|1\rangle_{at}$ are the ground and excited atomic states, respectively. Thus, $\hat{\sigma}_{i}|1\rangle_{at_i}=|0\rangle_{at_i}$ and $\hat{\sigma}_{i}^{\dagger}|0\rangle_{at_i}=|1\rangle_{at_i}$, where $i=1,2$ is the atomic index. Under resonance conditions, we set $\omega_{ph}=\omega_{at}=\omega$. Within the RWA, the atom--photon interaction Hamiltonian is given by:
	\begin{equation}
		\label{eq:H_int}
    		\hat{H}_{int}=g_{1}(\hat{a}^\dagger\hat{\sigma}_1+\hat{a}\hat{\sigma}_1^\dagger)+g_{2}(\hat{a}^\dagger\hat{\sigma}_2+\hat{a}\hat{\sigma}_2^\dagger) .
  	\end{equation} 
  
	The basis states of the system are given by:
	\begin{equation}
		\label{eq:Basis}
    		|n\rangle_{ph}|s_1\rangle_{at_{1}}|s_2\rangle_{at_{2}},
  	\end{equation}
  	where $|n\rangle_{ph}$ denotes the photon number state in the cavity, and $|s_1\rangle_{at_{1}}$ and $|s_2\rangle_{at_{2}}$ represent the states of the two atoms, respectively. For convenience, we abbreviate $|s_1\rangle_{at_{1}}|s_2\rangle_{at_{2}}$ as~$|s_1 s_2\rangle_{at}$.
  
  \subsection{Photon Exchange and Thermal Dephasing Channels}
  \label{subsec:DephasingChannels}

  Physically, pure dephasing corresponds to phase noise that suppresses coherence while leaving populations largely unchanged~\cite{Auffeves2009, Auffeves2010}. In our setting, thermal dephasing is used as an effective description of continuous QND monitoring. Unlike direct energy relaxation, it primarily damps the off-diagonal terms of the density matrix. It therefore degrades coherence without introducing a separate excitation manifold.
    
  The mapping from a continuous QND measurement to a pure dephasing Lindblad equation holds under two standard assumptions~\cite{Foroozani2016, Gross2018}: (i) unconditional (averaged) dynamics, where averaging over measurement outcomes removes the stochastic term, yielding a deterministic master equation and (ii) infinite-bandwidth (instantaneous) measurement, where the measurement is sufficiently fast and strong that the back-action reduces to pure dephasing at a constant rate $\gamma_{deph}$. Under these conditions, the continuous QND measurement of the atomic excitation numbers is described by the Lindblad operators $\hat{d}_i = |e_i\rangle\langle e_i|$.
  
  In the single-excitation subspace, the projectors are related to the Pauli $\hat{\sigma}_z$ operators as:
  \begin{equation}
  	\hat{d}_1 = |10\rangle\langle 10| = \frac{\hat{I} + \hat{\sigma}_z^{(1)}}{2}, \qquad
\hat{d}_2 = |01\rangle\langle 01| = \frac{\hat{I} + \hat{\sigma}_z^{(2)}}{2},
  \end{equation}
	where $\hat{\sigma}_z^{(1)} = \hat{\sigma}_z \otimes\hat{I}_2$, $\hat{\sigma}_z^{(2)} =\hat{I}_1\otimes \hat{\sigma}_z$, and $\hat{I}$ is the identity. Since the identity does not induce decoherence, the dephasing dynamics generated by $\hat{d}_i$ are equivalent to that generated by $\hat{\sigma}_z^{(i)}$ (up to an irrelevant global phase). This establishes the connection to dispersive QND readout in cavity QED.
	
	We note that the term ``thermal dephasing'' refers to the finite temperature of the photon bath, not to energy exchange in the dephasing process itself. The dephasing Lindblad operator $\hat{d}_i = |e_i\rangle\langle e_i|$ mimics the back-action of continuous QND measurement without population relaxation.
  
  We split the dissipative channels into two groups:
  \begin{itemize}
  	\item Photon exchange with the environment, represented by $\hat{a}$ and $\hat{a}^\dagger$:
  	\begin{subequations}
  		\label{eq:lindblad_photon_exchange}
  		\begin{align}
  			&\hat{\mathcal{L}}_{in} (\hat{\rho})=\gamma_{in}\left( \hat{a}^\dagger\hat{\rho} \hat{a} - \frac{1}{2}\left( \hat{a} \hat{a}^\dagger \hat{\rho} + \hat{\rho} \hat{a} \hat{a}^\dagger \right) \right),\label{eq:lindblad_photon_in}\\
  			&\hat{\mathcal{L}}_{out}(\hat{\rho})=\gamma_{out}\left(\hat{a}\hat{\rho}\hat{a}^\dagger - \frac{1}{2}\left(\hat{a}^\dagger \hat{a}\hat{\rho} + \hat{\rho} \hat{a}^\dagger \hat{a} \right) \right).\label{eq:lindblad_photon_out}
  		\end{align}
  	\end{subequations}
  	\item Thermal dephasing from QND monitoring, represented by projectors onto the atomic excitation subspaces $\hat{d}_1$ and $\hat{d}_2$ and their conjugate operators: 
  	\begin{subequations}
  		\label{eq:lindblad_thermal_dephasing}
  		\begin{align}
        		&\hat{\mathcal{L}}_{deph,1}(\hat{\rho}) =\gamma_{deph,1}\left( \hat{d}_{1}\hat{\rho} \hat{d}_{1}^\dagger - \frac{1}{2}\left( \hat{d}_{1}^\dagger \hat{d}_{1} \hat{\rho} + \hat{\rho} \hat{d}_{1}^\dagger \hat{d}_{1} \right) \right),\label{eq:lindblad_thermal_dephasing_at1}\\
        		&\hat{\mathcal{L}}_{deph,2}(\hat{\rho}) =\gamma_{deph,2}\left( \hat{d}_{2}\hat{\rho} \hat{d}_{2}^\dagger - \frac{1}{2}\left( \hat{d}_{2}^\dagger \hat{d}_{2} \hat{\rho} + \hat{\rho} \hat{d}_{2}^\dagger \hat{d}_{2} \right) \right),\label{eq:lindblad_thermal_dephasing_at2}
  		\end{align}
  	\end{subequations}
  	where $\hat{d}_1=\hat{\sigma}_1^\dagger \hat{\sigma}_1\otimes \hat{I}_1$ and $\hat{d}_2=\hat{I}_2 \otimes\hat{\sigma}_2^\dagger\hat{\sigma}_2$.
  \end{itemize}
  
  Thus, Equation \eqref{eq:QME} can be rewritten as:
  \begin{equation}
  	\label{eq:NewQME}
  	i\hbar\dot{\hat{\rho}}=[\hat{H},\hat{\rho}]+i\left(\hat{\mathcal{L}}_{in}(\hat{\rho})+\hat{\mathcal{L}}_{out} (\hat{\rho})+\hat{\mathcal{L}}_{deph,1}(\hat{\rho})+\hat{\mathcal{L}}_{deph,2}(\hat{\rho})\right).
  \end{equation}
  Thus, $\hat{\mathcal{L}}(\hat{\rho})=\hat{\mathcal{L}}_{in}(\hat{\rho})+\hat{\mathcal{L}}_{out} (\hat{\rho})+\hat{\mathcal{L}}_{deph,1}(\hat{\rho})+\hat{\mathcal{L}}_{deph,2}(\hat{\rho})$.
  
  In our model, $\gamma_{deph,1}$ and $\gamma_{deph,2}$ are treated as independent parameters. This allows us to study the robustness of Zeno/anti-Zeno effects against asymmetries that naturally appear in realistic experiments. For example, when the two atoms are placed at different positions of the cavity standing wave, their coupling strengths $g_i$ differ, leading to $\gamma_{deph,i} \propto g_i^2$ for dispersive QND measurements. Asymmetric dephasing also appears in individually addressed measurement setups or due to experimental imperfections. The symmetric case ($\gamma_{deph,1}=\gamma_{deph,2}$) is recovered as a special limit.
  
  We explicitly state the main approximations: (i) Markovian bath with flat spectrum (justified by $\tau_{bath} \ll 1/g$); (ii) zero detuning ($\omega_{ph} = \omega_{at}$), with off-resonant effects left for future work; (iii) single-excitation subspace, ensured by initial condition and global energy constraint; (iv) RWA applied to the Hamiltonian, with no additional secular approximation in the master equation.
  
	\section{Numerical Methods}
	\label{sec:Method}

	To solve the Lindblad master equation in Equation \eqref{eq:NewQME}, we employ explicit time-stepping methods. Because the Hilbert space dimension in our simulations is relatively small, these methods are both convenient and computationally inexpensive. We implement two approaches: a first-order Euler-type method as a baseline and the fourth-order Runge--Kutta (RK4) method for improved accuracy.
	
	\subsection{Euler Method}
	\label{subsec:Eulermethod}

	As a simple baseline, we use a first-order update that separates the coherent and dissipative dynamics. The coherent evolution over a time step $\Delta t$ is generated by the Hamiltonian $\hat{H}$, while the dissipative correction is provided by the Lindblad superoperator~$\hat{\mathcal{L}}(\hat{\rho})$:
\begin{equation}
	\label{eq:Eulermethod}
	\hat{\rho}(t+\Delta t) =\tilde{\hat{\rho}}(t+\Delta t)+ \frac{1}{\hbar} \hat{\mathcal{L}}(\tilde{\hat{\rho}}(t+\Delta t))\Delta t,
\end{equation}
where
\begin{equation}
	\label{eq:Intermediate}
	\tilde{\hat{\rho}}(t+\Delta t)=\exp\left(\frac{-i\hat{H}\Delta t}{\hbar}\right)\hat{\rho}(t)\exp\left(\frac{i\hat{H}\Delta t}{\hbar}\right)
\end{equation}

	\subsection{Runge--Kutta Method}
	\label{subsec:RungeKuttaMethod}
	
	For higher accuracy, we employ the classical RK4 method. Equation \eqref{eq:NewQME} can be rewritten as:
	\vspace{3pt}\begin{equation}
    		\label{eq:rho_dot_Lrho}
    		 \dot{\hat{\rho}}=\hat{\mathcal{L}}_{entire}(\hat{\rho})=-\frac{i}{\hbar}[\hat{H},\hat{\rho}]+\frac{1}{\hbar}\hat{\mathcal{L}}(\hat{\rho}),
	\end{equation}
	where $\hat{\mathcal{L}}_{entire}(\hat{\rho})$ denotes the full Liouvillian superoperator, which includes both the Hamiltonian commutator term and all Lindblad dissipators. Equation \eqref{eq:rho_dot_Lrho} is a first-order ordinary differential equation with $\hat{\rho}(t)$ as a matrix-valued unknown. Therefore, standard ODE solvers apply without modification. The RK4 method computes four intermediate slopes by evaluating $\hat{\mathcal{L}}_{entire}$ at shifted arguments:
	\begin{equation}
		\label{eq:rk4_rho}
      	\hat{\rho}(t+\Delta t) = \hat{\rho}(t) + \frac{\Delta t}{6}\left(k_1+2k_2+2k_3+k_4\right),
	\end{equation}
	where 
    \begin{subequations}
        \label{eq:rk4_rho_steps}
        \begin{align}
            k_1&=\hat{\mathcal{L}}_{entire}(\hat{\rho}(t)),\label{eq:rk4_rho_step1}\\
            k_2&=\hat{\mathcal{L}}_{entire}\left(\hat{\rho}(t) + \frac{\Delta t}{2}\,k_1\right),\label{eq:rk4_rho_step2}\\
            k_3&= \hat{\mathcal{L}}_{entire}\left(\hat{\rho}(t) + \frac{\Delta t}{2}\,k_2\right),\label{eq:rk4_rho_step3}\\
            k_4&= \hat{\mathcal{L}}_{entire}\left(\hat{\rho}(t) + \Delta t\,k_3\right).\label{eq:rk4_rho_step4}
        \end{align}
    \end{subequations}

	This method has a local truncation error of order $\mathcal{O}(\Delta t^{5})$ and a global error of order $\mathcal{O}(\Delta t^{4})$. The RK4 method thus offers a convenient accuracy--cost trade-off: one step requires four evaluations of $\hat{\mathcal{L}}_{entire}$, but the improved accuracy often permits a larger $\Delta t$ compared to Euler updates.

	\subsection{Choice of Time Step and Numerical Stability}
	\label{subsec:Stability}

	In optical cavity QED experiments, the time step is often limited by the measurement time, which is approximately $1\mu\text{s}$~\cite{Hennrich2000}. In practice, the time step must be chosen sufficiently small to resolve the fastest dynamical scale in the problem. This scale is determined by the largest coherent frequency and the largest dissipative rate.
	
	While the exact Lindblad evolution preserves $\hat{\rho}\succeq 0$ (positive semidefiniteness), \mbox{$\mathrm{Tr}\hat{\rho}=1$}, and $\hat{\rho}=\hat{\rho}^\dagger$ (Hermiticity), explicit discretizations may slightly violate these constraints if $\Delta t$ is too large. Numerical round-off errors can also accumulate over long integration times. In our simulations, we therefore
	\begin{itemize}
    		\item Choose $\Delta t$ sufficiently small to resolve all relevant timescales;
    		\item Monitor $\mathrm{Tr}\hat{\rho}$ and Hermiticity during the run;
    		\item If needed, apply small post-step corrections such as $\hat{\rho} \leftarrow (\hat{\rho}+\hat{\rho}^\dagger)/2$ and $\hat{\rho} \leftarrow \hat{\rho}/\mathrm{Tr}\hat{\rho}$ to suppress the accumulation of floating-point errors.
	\end{itemize}
	
	In our converged simulations, the magnitude of these post-step corrections is negligibly small compared to the physical observables (typically well below $10^{-9}$). The exact numerical values depend on the time step and parameters, but they remain far below any physically relevant scale. If a noticeable trace drift or a significant non-Hermitian contribution appears, this should be treated as a warning sign. In practice, such behavior means that the chosen time step $\Delta t$ is too large, the integrator is being pushed outside its reliable range, or the numerical tolerance must be tightened. In that case, the integration parameters must be refined.
	
	All simulations are performed using double-precision (64-bit) floating-point arithmetic. The trace of the density matrix remains within $1 \pm 10^{-12}$ throughout the evolution, confirming trace conservation to machine precision. Positivity is also monitored an\mbox{d m}aintained.

	For systems with a large number of states, where explicit density-matrix integration becomes prohibitive, alternative methods can be used. These include the Monte Carlo wave function method~\cite{Dum1992} and the quantum trajectory method~\cite{Dalibard1992, Klaus1993}. In the present work, however, the Hilbert space dimension is small enough that direct integration of Equat\mbox{ion \eqref{eq:rho_dot_Lrho}} is computationally feasible.

\section{Results}
  \label{sec:Results}
  
  In cavity QED experiments, typical parameters are as follows: cavity frequency: $1$--$10$ GHz, atom--cavity coupling strength: on the order of MHz, photon decay rate: $0.1$--$10$ MHz. These ranges are based on the experimental literature~\cite{Fink2009, Bosman2017, Zhou2024}. Our simulations use parameters within these experimentally accessible regimes.

  For two atoms with coupling strengths $g_1$ and $g_2$, the target dark state in the one-excitation sector is proportional to $|n\rangle_{ph}(g_1|01\rangle-g_2|10\rangle)_{at}$. We analyze its stability under thermal dephasing and identify parameter regions that maximize the retained dark-state weight. We also study the evolution from initial conditions that are not dark states.

  The photon exchange operators $\hat{a}$ and $\hat{a}^\dagger$ are associated with the outflow and inflow rates $\gamma_{out}$ and $\gamma_{in}$, respectively~\cite{Afanasyev2022}. In the dissipative regime considered here, we have $\gamma_{out}>\gamma_{in}$. For a fixed mode frequency, one can introduce the effective thermal parameter~\cite{Kulagin2019}:
 \vspace{3pt} \begin{equation}
      \label{eq:RatioPhotonExchange}
      \mu_{ph}=\frac{\gamma_{in}}{\gamma_{out}}=\exp\left(-\frac{\hbar\omega}{KT_{\omega}}\right),
  \end{equation}
  where $K$ is the Boltzmann constant and $T_{\omega}$ is the cavity temperature associated with mode~$\omega$.

  For the thermal dephasing operators $\hat{d}_1$ and $\hat{d}_2$, we will use $k$ to represent the ratio between their corresponding dephasing rates:
  \begin{equation}
  	  \label{eq:RatioDephashing}
      k=\frac{\gamma_{deph,1}}{\gamma_{deph,2}},
  \end{equation}
  where $\gamma_{deph,1}$ and $\gamma_{deph,2}$ quantify the effective observation strength on each atom. Larger values correspond to stronger dephasing, i.e., more frequent monitoring. The case $k=1$ corresponds to symmetric observation, while $k \neq 1$ corresponds to asymmetric observation.
  
  \subsection{Atomic Dephasing Simulation of Two-Level Atomic System}
  \label{subsec:Results_two_level}
  
  \begin{figure}[H]
  \hspace{-14pt} 	\includegraphics[width=1.\textwidth]{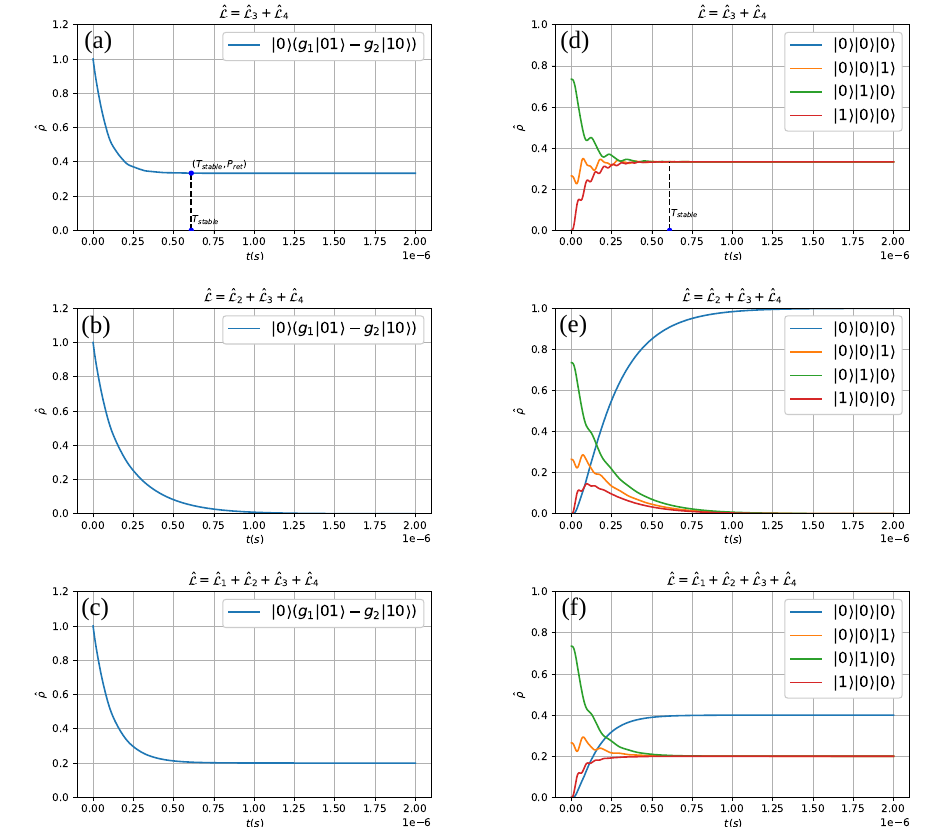}
    \caption{(Online color) {\it Influence of cavity photon exchange on dark-state retention under thermal dephasing.} Panels (\textbf{a}--\textbf{c}) show the dark-state population under the different conditions, and panels (\textbf{d}--\textbf{f}) show the populations of all states. The operators $\hat{\mathcal{L}}_1$, $\hat{\mathcal{L}}_2$, $\hat{\mathcal{L}}_3$ and $\hat{\mathcal{L}}_4$ correspond to $\hat{\mathcal{L}}_{in}$, $\hat{\mathcal{L}}_{out}$, $\hat{\mathcal{L}}_{deph,1}$ and $\hat{\mathcal{L}}_{deph,2}$, respectively. Parameters: $\omega = 1\,\mathrm{GHz}$, $g_1 = 30\,\mathrm{MHz}$, $g_2 = 50\,\mathrm{MHz}$, $\gamma_{out}=20\,\mathrm{MHz}$, $\gamma_{in}=10\,\mathrm{MHz}$, and $\gamma_{deph,1} = \gamma_{deph,2} = 20\,\mathrm{MHz}$.}
 \label{fig:InfluencePhotonExchange}
  \end{figure}

  Based on the discussion in Section~\ref{subsec:DephasingChannels}, we numerically verify three effects: the role of cavity photon exchange in dark-state preservation, the transition between the anti-Zeno and Zeno regimes as the dephasing strength varies, and the dependence of optimal preservation on field factors.

  Figure \ref{fig:InfluencePhotonExchange} compares three dissipative scenarios from left to right. In panel (a), only dephasing is present, i.e., $\hat{\mathcal{L}}=\hat{\mathcal{L}}_{deph,1}+\hat{\mathcal{L}}_{deph,2}$. The dark superposition loses coherence and decays into the $|01\rangle_{at}/|10\rangle_{at}$ components. In panel (b), photon outflow is added, so $\hat{\mathcal{L}}=\hat{\mathcal{L}}_{out}+\hat{\mathcal{L}}_{deph,1}+\hat{\mathcal{L}}_{deph,2}$. Consequently, dark-state retention is reduced further. In panel (c), both outflow and inflow are present, i.e., $\hat{\mathcal{L}}=\hat{\mathcal{L}}_{out}+\hat{\mathcal{L}}_{in}+\hat{\mathcal{L}}_{deph,1}+\hat{\mathcal{L}}_{deph,2}$. As a result, retention improves relative to the outflow-only case. The corresponding population trajectories in panels (d)--(f) show the same trend in greater detail. Without replenishment, photon leakage accelerates the loss of dark-state population. When replenishment is present, the dark component remains finite for longer times, which indicates partial stabilization against thermal dephasing. We define the stabilization time $T_{stab}$ by the criterion that for each basis state $|a\rangle$ and for all $t > T_{stab}$, the condition $|P_{|a\rangle}(t) - P_{|a\rangle}(T_{stab})| < 10^{-8}$ holds. The system is integrated up to $t_{max} = 2 T_{stab}$. Extending the integration further changes $P_{ret}$ by less than $10^{-8}$, confirming a true steady state. The dark-state retention metric is then defined as $P_{ret}=P_{|dark\rangle}(T_{stab})$. Because $T_{stab}$ spans a broad range across parameters, we report normalized heatmaps of the stabilization time and retention metrics.
  
  	\begin{figure}[H]
   	\includegraphics[width=0.99\textwidth]{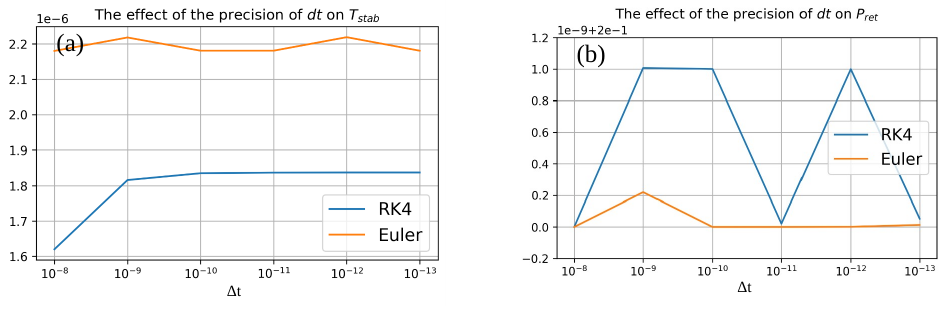}
    \caption{(Online color) {\it Convergence test for $T_{stab}$ and $P_{ret}$ as functions of the time step $\Delta t$.} Panel (\textbf{a}) $T_{stab}$ and panel (\textbf{b}) $P_{ret}$ as functions of the time step $\Delta t$. The RK4 method converges faster than the Euler method. Parameters: $g_1=30$ MHz, $g_2=50$ MHz, $\gamma_{out}=20$ MHz, $\gamma_{in}=10$ MHz, $\gamma_{deph,1}=\gamma_{deph,2}=20$ MHz.}
 \label{fig2}
  \end{figure}
  
   Based on a systematic convergence test (see Figure \ref{fig2}), we adopt a fixed time step $\Delta t = 10^{-12}$ s for all production simulations. This value is well within the converged regime and resolves the fastest coherent dynamics (minimum period $5$ ns) with at least 5000 steps per period.
   
\begin{figure}[H]
   	\includegraphics[width=.9\textwidth]{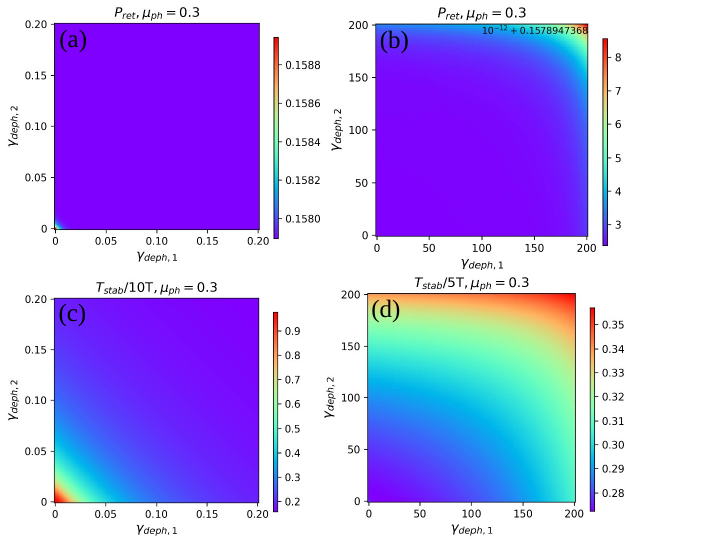}
    \caption{(Online color) {\it Heatmaps of dark-state retention and stabilization time versus dephasing rates $\gamma_{deph,1}$ and $\gamma_{deph,2}$.} Panels (\textbf{a},\textbf{b}) show $P_{ret}$, while panels (\textbf{c},\textbf{d}) show $T_{stab}$. The coupling strengths $\gamma_{deph,1}$ and $\gamma_{deph,2}$ are both given in $\mathrm{MHz}$. Parameters: $\omega = 1\,\mathrm{GHz}$, $g_1 = 30\,\mathrm{MHz}$, $g_2 = 30\,\mathrm{MHz}$, $\gamma_{out}=10\,\mathrm{MHz}$, $\gamma_{in}=3\,\mathrm{MHz}$.}
 	\label{fig:Heatmap1}
  \end{figure}

    Numerous studies have indicated that increasing the observation intensity at low levels leads to the anti-Zeno effect, while at higher levels, the Zeno effect occurs~\cite{Davidsson2023, Chaudhry2014, Kofman2000}. Therefore, we investigate whether dark states exhibit similar behavior. We vary the thermal dephasing intensity over two distinct ranges. A larger dephasing rate $\gamma_{deph}$ corresponds to stronger observation-induced phase randomization. Since the response spans several orders of magnitude, we split the scan into two intervals: $[0,0.2]$ (MHz) (panels (a) and (c) of Figure \ref{fig:Heatmap1}) and $[0,200]$ (MHz) (panels (b) and (d) of Figure \ref{fig:Heatmap1}). At low observation intensity, increasing the dephasing rate slightly reduces $T_{stab}$ (faster relaxation) and also decreases $P_{ret}$. At higher intensity, the trend reverses: $T_{stab}$ increases and $P_{ret}$ grows slowly, indicating the onset of Zeno-like freezing. Therefore, the model exhibits a crossover from anti-Zeno behavior (accelerated dynamics) at weak dephasing to Zeno behavior (slowed dynamics) at stronger dephasing. Importantly, even in the strong-dephasing regime, the dark-state component remains finite, confirming its robustness.
  
\begin{figure}[H]
   \hspace{-3pt}	\includegraphics[width=0.9\textwidth]{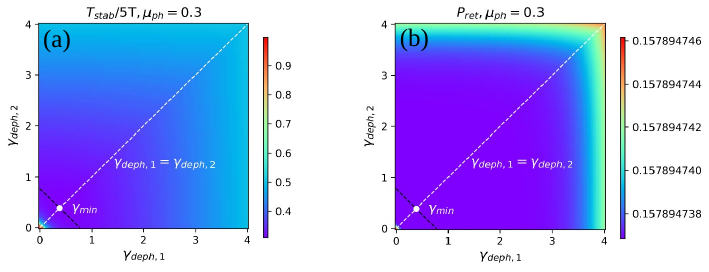}
    \caption{(Online color) {\it Trends of $T_{stab}$ and $P_{ret}$.} In panels (\textbf{a}) and (\textbf{b}), the black dashed curve marks the anti-Zeno to Zeno boundary. Its intersection with the white diagonal gives the point where $\gamma_{deph,1}=\gamma_{deph,2}=\gamma_{\min}$. The dephasing rates $\gamma_{deph,1}$ and $\gamma_{deph,2}$ are given in $\mathrm{MHz}$. Parameters: $\mu_{ph}=0.3$, $\gamma_{out}=10\,\mathrm{MHz}$, $\gamma_{in}=3\,\mathrm{MHz}$. $g_1 = 2\,\mathrm{MHz}$, $g_2 = 2\,\mathrm{MHz}$.}
	\label{fig:Heatmap2}
  \end{figure}

\begin{figure}[H]
   	\includegraphics[width=0.99\textwidth]{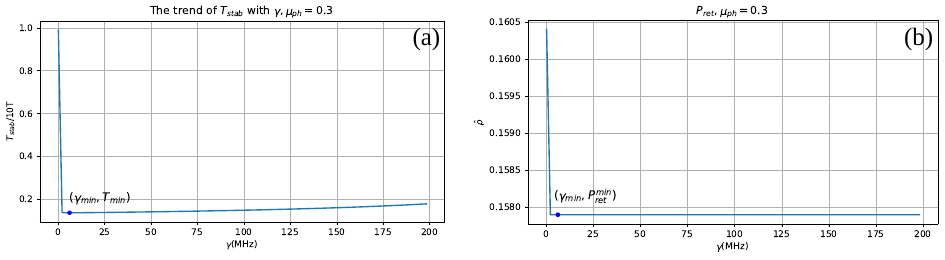}
    \caption{(Online color) {\it Cross-sections along the line $\gamma_{deph,1}=\gamma_{deph,2}$.} Panel (\textbf{a}) shows $T_{stab}$ and the location of $\gamma_{\min}$, while panel (\textbf{b}) shows $P_{ret}$. The data illustrate the strong robustness of dark states: after a shallow initial decrease, $P_{ret}$ rises slowly with increasing dephasing, consistent with the anti-Zeno to Zeno crossover. Parameters: $\mu_{ph}=0.3$, $\gamma_{out}=10\,\mathrm{MHz}$, $\gamma_{in}=3\,\mathrm{MHz}$.$g_1 = 2\,\mathrm{MHz}$, $g_2 = 2\,\mathrm{MHz}$.}
 \label{fig:Cross-sections}
  \end{figure}

  We first analyze the low-dephasing regime in Figure \ref{fig:Heatmap2}. The dark central region and brighter edges indicate a non-monotonic response: as $\gamma_{deph,1}$ and $\gamma_{deph,2}$ increase, the system first enters an anti-Zeno regime and then crosses into a Zeno regime. To reduce complexity, we first set $g_1=g_2$. Along the diagonal line $\gamma_{deph,1} = \gamma_{deph,2}$, the stabilization time exhibits a distinct minimum, which defines $\gamma_{min}$. We then explore how the shortest equilibration time $T_{min}$, achieved when $\gamma_{deph,1}=\gamma_{deph,2}=\gamma_{\min}$, varies with the coupling strength $g$ under the symmetric condition $g_1 = g_2 = g$. Figure \ref{fig:Cross-sections} summarizes the behavior along this diagonal. For $\gamma_{deph,1}=\gamma_{deph,2}<\gamma_{\min}$, $T_{stab}$ decreases (anti-Zeno regime) and $P_{ret}$ decreases slightly. For $\gamma_{deph,1}=\gamma_{deph,2}>\gamma_{\min}$, $T_{stab}$ increases (Zeno regime) and $P_{ret}$ grows slowly.
  
  \begin{figure}[H]
   	\hspace{-3pt}\includegraphics[width=.5\textwidth]{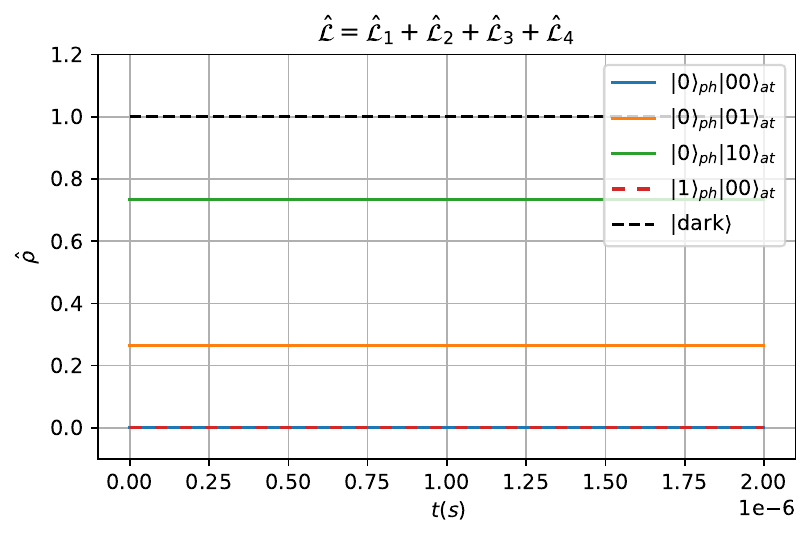}
    \caption{(Online color) {\it Influence of cavity photon exchange on dark-state retention under thermal dephasing with $\gamma_{deph,1}=\gamma_{deph,2}=0$.} The colored lines and black dashed lines correspond to the changing trends of quantum states and dark states $|s\rangle$ over time, respectively. The operators $\hat{\mathcal{L}}_1$, $\hat{\mathcal{L}}_2$, $\hat{\mathcal{L}}_3$ and $\hat{\mathcal{L}}_4$ correspond to $\hat{\mathcal{L}}_{in}$, $\hat{\mathcal{L}}_{out}$, $\hat{\mathcal{L}}_{deph,1}$ and $\hat{\mathcal{L}}_{deph,2}$, respectively. Parameters: $\omega = 1\,\mathrm{GHz}$, $g_1 = 30\,\mathrm{MHz}$, $g_2 = 50\,\mathrm{MHz}$, $\gamma_{out}=20\,\mathrm{MHz}$, $\gamma_{in}=10\,\mathrm{MHz}$, and $\gamma_{deph,1} = \gamma_{deph,2} = 0$.}
 \label{fig6}
  \end{figure}
  
\begin{figure}[H]
   	\includegraphics[width=0.99\textwidth]{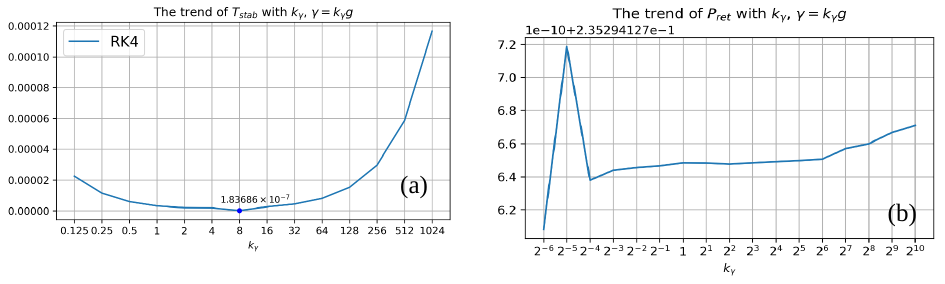}
    \caption{(Online color) {\it Non-monotonic behavior of $T_{stab}$ and frequency-independent $P_{ret}$ as functions of $k_\gamma=\gamma_{deph}/g$.} Panel (\textbf{a}) $T_{stab}$ vs. $k_\gamma = \gamma_{deph}/g$, showing a U-shaped non-monotonic curve and exponential growth for $k_\gamma>128$. Panel (\textbf{b}) $P_{ret}$ vs. $k_\gamma$, remaining constant at $\approx 0.235$ over five orders of magnitude. Parameters: $\mu_{ph}=0.8$, $g_1=g_2=10$ MHz, $\gamma_{out}=12.5$ MHz, $\gamma_{in}=10$ MHz.}
 \label{fig7}
  \end{figure}

  Figure \ref{fig6} verifies the expected limiting behavior: for $\gamma_{deph}=0$, the dark-state population is strictly constant, confirming that the dark state is perfectly stable in the absence of measurement. For the representative parameters shown in Figure \ref{fig7}, we identify the following based on $k_\gamma=\gamma_{deph}/g$:
	\begin{itemize}
		\item Anti-Zeno regime: $k_\gamma \lesssim 8$ (accelerated relaxation).
		\item Zeno regime: $k_\gamma \gtrsim 8$ (slowed relaxation), with exponential growth for $k_\gamma > 128$.
	\end{itemize}
	
	The transition occurs at $k_\gamma \approx 8$, consistent with the Kofman--Kurizki theory~\cite{Kofman2000}. Classical saturation would not produce such non-monotonic behavior.
  
\begin{figure}[H]
   	\includegraphics[width=.5\textwidth]{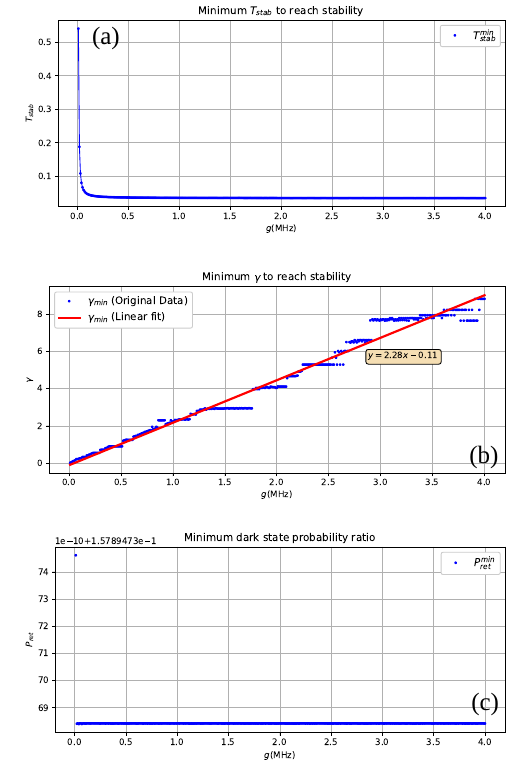}
    \caption{(Online color) {\it Field-factor dependence of optimal quantities under symmetric dephasing \mbox{$\gamma_{deph,1}=\gamma_{deph,2}$}.} Panel (\textbf{a}) shows the minimum stabilization time $T_{stab}^{min}$, panel (\textbf{b}) shows the corresponding optimal dephasing rate $\gamma_{\min}$, and panel (\textbf{c}) shows the minimum dark-state retention~$P_{ret}^{min}$.}
    \label{fig:min_panels}
  \end{figure}

  Figure \ref{fig:min_panels}a shows the shortest evolution time achievable for $g_1=g_2=g$ by adjusting $\gamma_{deph,1}$ and $\gamma_{deph,2}$ (with $\gamma_{deph,1}=\gamma_{deph,2}=\gamma_{\min}$). Blue dots represent numerical data, and red dots show the fitted curve. Panel (b) shows that the optimal dephasing rate $\gamma_{\min}$ is positively correlated with the coupling strength $g$. Panel (c) shows the minimum retained dark-state probability. For $g_1=g_2=g$, this probability remains close to $0.1579$, with only a weak dependence on $g$. Small fluctuations are within numerical uncertainty. Overall, these scans confirm that dark states are resilient to thermal dephasing. For $g_1=g_2=g$, there exists an optimal dephasing rate $\gamma_{\min}$ (which is dependent on $g$) that minimizes the stabilization time. Near this point, the dark-state survival probability also reaches a local minimum. When the initial state is a dark state and the cavity exchanges photons with the environment, $\gamma<\gamma_{\min}$ corresponds to anti-Zeno behavior, characterized by a slight decrease in the dark-state population. In contrast, $\gamma>\gamma_{\min}$ corresponds to Zeno behavior, with a gradual recovery of dark-state retention.
  
\begin{figure}[H]
   	\includegraphics[width=0.9\textwidth]{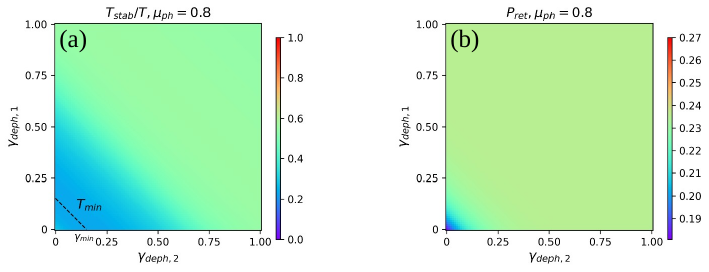}
    \caption{(Online color) {\it Heatmaps of $T_{stab}$ and $P_{ret}$ versus thermal dephasing intensity for the non-dark initial state $|0\rangle_{ph}|01\rangle_{at}$.} Panel (\textbf{a}) shows the heatmap of $T_{stab}$, and panel (\textbf{b}) shows the heatmap of $P_{ret}$. The dephasing rates $\gamma_{deph,1}$ and $\gamma_{deph,2}$ are given in $\mathrm{MHz}$. Parameters: $\mu_{ph}=0.8$, $\gamma_{out}=10\,\mathrm{MHz}$, $\gamma_{in}=8\,\mathrm{MHz}$.$g_1 = 30\,\mathrm{MHz}$, $g_2 = 70\,\mathrm{MHz}$.}
 \label{fig:Heatmap3}
  \end{figure}

     Next, we consider a more general case where the initial state is not a dark state. For example, we take the state with no photons in the cavity, one atom in the ground state, and the other in the excited state: $|0\rangle_{ph}|01\rangle_{at}$. The map in Figure \ref{fig:Heatmap3}a has a darker center and lighter edges, indicating the non-monotonic dependence of the stabilization time on dephasing rate. Even from a non-dark initial condition, a finite dark-state component emerges during the evolution. The near-symmetry of the map suggests comparable sensitivity to $\gamma_{deph,1}$ and $\gamma_{deph,2}$. This non-dark initial state also exhibits a clear Zeno/anti-Zeno crossover. As the dephasing rate $\gamma_{deph}$ increases, the stabilization time $T_{stab}$ first decreases (anti-Zeno regime) and then increases (Zeno regime), with a minimum at an intermediate $\gamma_{deph}$. Moreover, the steady-state dark-state population $P_{ret}$ becomes independent of $\gamma_{deph}$ once the measurement is turned on, similar to the dark-state case. These observations indicate that the Zeno/anti-Zeno phenomenon is not specific to the ideal dark state but is a robust feature of the two-atom cavity QED system under continuous dephasing.

\begin{figure}[H]
     \hspace{-5pt}	\includegraphics[width=0.9\textwidth]{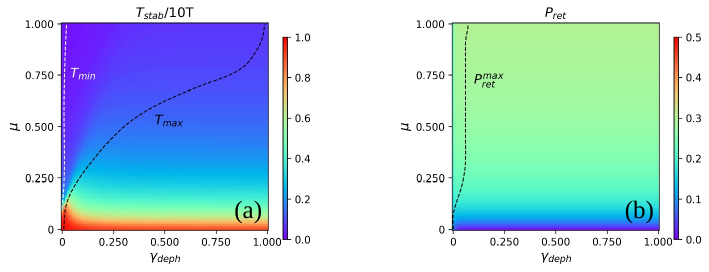}
    \caption{(Online color) {\it Effects of thermal dephasing intensity and temperature on $T_{stab}$ (panel (\textbf{a})) and $P_{ret}$ (panel (\textbf{b})) under symmetric measurement conditions $\gamma_{deph,1}=\gamma_{deph,2}$ (initial state is not a dark state).} The dephasing rates $\gamma_{deph,1}$ and $\gamma_{deph,2}$ are given in $\mathrm{MHz}$. Parameters: $g_1 = 30\,\mathrm{MHz}$, $g_2 = 70\,\mathrm{MHz}$. A higher temperature increases $\mu_{ph} = \gamma_{in}/\gamma_{out}$, which raises the photon inflow rate $\gamma_{in}$. Under the global energy constraint, this reduces the net photon loss and drives the system toward a steady state with larger overlap with the dark state, thereby enhancing dark-state retention.}
 \label{fig:Heatmap4}
  \end{figure}

  Under symmetric measurement conditions ($\gamma_{deph,1}=\gamma_{deph,2}=\gamma_{deph}$), we vary $\mu_{ph}$ and the dephasing rate simultaneously. The retention metric $P_{ret}$ is positively correlated with $\mu_{ph}$. The brighter upper-right region in Figure \ref{fig:Heatmap4}a indicates enhanced dark-state preservation at higher temperatures and stronger dephasing. The behavior of $T_{stab}$ in Figure~\ref{fig:Heatmap4}b is qualitatively similar: for fixed $g_1/g_2$, increasing $\gamma_{deph}$ first accelerates and then slows down the dynamics, while a larger $\mu_{ph}$ mainly amplifies this contrast. The dashed ridge separates the anti-Zeno- and Zeno-dominated regions.
  
\begin{figure}[H]
   \hspace{-4pt}	\includegraphics[width=0.9\textwidth]{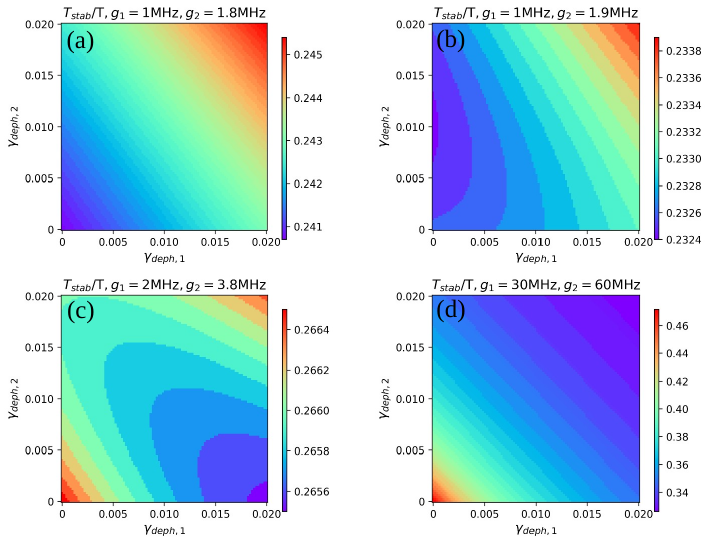}
    \caption{(Online color) {\it Representative heatmap types for different values of $\kappa$ in the low-dephasing region $(\gamma_{deph,1},\gamma_{deph,2})$ at $\mu_{ph}=0.8$.} Panel (\textbf{a}) corresponds to $\kappa<\kappa_0$, panels (\textbf{b},\textbf{c}) correspond to $\kappa\approx \kappa_0$, and panel (\textbf{d}) corresponds to $\kappa>\kappa_0$. The dephasing rates $\gamma_{deph,1}$ and $\gamma_{deph,2}$ are given in $\mathrm{MHz}$. Parameters: $\gamma_{out}=10\,\mathrm{MHz}$, $\gamma_{in}=8\,\mathrm{MHz}$.}
 \label{fig:Heatmap5}
  \end{figure}

   Since the temperature mainly rescales the dephasing response, we next fix $\mu_{ph} = 0.8$ and analyze asymmetric measurement conditions, i.e., $\gamma_{deph,1} \neq \gamma_{deph,2}$. The transition pattern between anti-Zeno and Zeno behavior falls into three classes, which are controlled by the coupling ratio $\kappa=g_2/g_1$. We parameterize the couplings as $g_2=\kappa\,g_1$. For each value of $g_1$, there exists a threshold $\kappa_0$ that separates different dynamical regimes:
\begin{itemize}
	\item For $\kappa<\kappa_0$, the behavior along the diagonal $\gamma_{deph,1}=\gamma_{deph,2}$ initially increases (Zeno-first pattern; Figure \ref{fig:Heatmap5}a);
	\item For $\kappa\approx \kappa_0$, the map exhibits a transitional saddle-like structure (Figure \ref{fig:Heatmap5}b,c);
	\item For $\kappa>\kappa_0$, the behavior along the diagonal first decreases and then increases (anti-Zeno to Zeno pattern; Figure \ref{fig:Heatmap5}d).
\end{itemize}

\begin{figure}[H]
   	\includegraphics[width=0.7\textwidth]{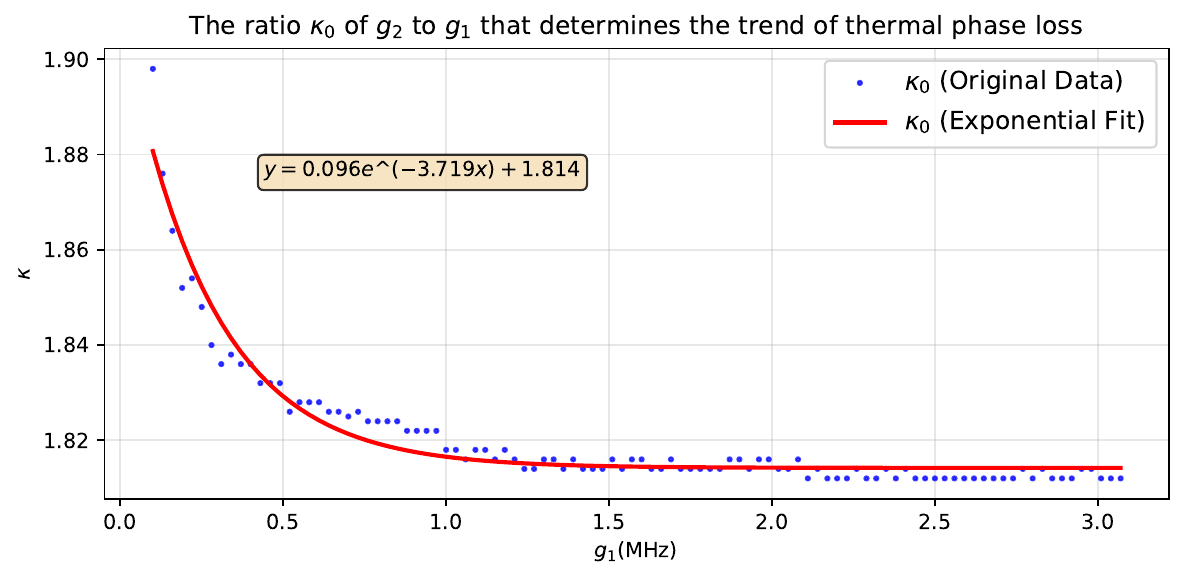}
    \caption{(Online color) {\it The value of $\kappa_0$ decreases approximately exponentially with increasing $g_1$.} Parameters: $\mu_{ph}=0.8$, $\gamma_{out}=10\,\mathrm{MHz}$, $\gamma_{in}=8\,\mathrm{MHz}$.} 
    \label{fig:k0}
  \end{figure}

	Thus, the coupling ratio $\kappa$ determines whether increasing dephasing drives the system from anti-Zeno to Zeno behavior or in the opposite order. The critical value $\kappa_0$ (Figure~\ref{fig:k0}) is approximately $1.8$ in our scans and decreases slowly as $g_1$ increases.

  \subsection{Reduced Three-State Illustration: Populations and Fidelity Under Dephasing}
  \label{subsubsec:reduced_three_state_results}
  
  To gain additional insight into how thermal dephasing destroys the dark-state property of the singlet-like state, it is convenient to restrict the dynamics to the one-excitation manifold of the TC model. In this subsection, we quantify the overlap with the target dark state by the fidelity $F(\hat{\rho},\hat{\sigma}_D)$. In the simplest truncation, we consider the three basis states: $\{|1\rangle|00\rangle$, $|0\rangle|01\rangle$, $|0\rangle|10\rangle\}$. These correspond to either one photon in the cavity with both atoms in the ground state, or no photon with exactly one atom excited.

  In this reduced picture, the dark state corresponds to the antisymmetric atomic superposition with no photon in the cavity:
  \begin{equation}
    |\Psi_D\rangle = |0\rangle_{ph}\otimes \frac{g_1|01\rangle_{at}-g_2|10\rangle_{at}}{\sqrt{|g_1|^2+|g_2|^2}}.
    \label{eq:dark_state_reduced}
  \end{equation}
  Pure dephasing does not change the populations in the energy basis. However, it suppresses coherences between $|01\rangle_{at}$ and $|10\rangle_{at}$, thereby turning $|\Psi_D\rangle$ into an incoherent mixture and reducing its overlap with the target dark state.
  
  To quantify the loss of ``darkness'', we compute the fidelity between the instantaneous density matrix $\hat{\rho}$ and the target state $\hat{\sigma}_D = |\Psi_D\rangle\langle\Psi_D|$:
  \begin{equation}
        F(\hat{\rho},\hat{\sigma}_D)=\left(\mathrm{Tr}\sqrt{\sqrt{\hat{\rho}}\,\hat{\sigma}_D\,\sqrt{\hat{\rho}}}\right)^2=\langle\Psi_D|\hat{\rho}|\Psi_D\rangle.
  \end{equation}
  Since $\hat{\sigma}_D$ is a projector onto a pure state, the Uhlmann fidelity reduces in the present case to the expectation value of $\hat{\rho}$ in the target dark state. The quantity $F$ therefore lies between $0$ and $1$: $F=1$ corresponds to an exactly preserved dark state, whereas smaller values measure the loss of overlap caused by dephasing and photon exchange. The same notation is used consistently in the text, equations, and figure captions below.
  
  In the following simulations, we test three hypotheses:
  \begin{itemize}
  	\item Dephasing alone suppresses the coherence of the dark superposition.
  	\item Photon exchange channels modify this suppression and can partially recover the dark-state population.
  	\item When the initial state is dark, cavity photon exchange can stabilize a finite dark-state component even in the presence of dephasing.
  \end{itemize}
  
  Before presenting the full parameter scans, we show a compact numerical illustration within the reduced three-state basis $\{|1\rangle|00\rangle$, $|0\rangle|01\rangle$, $|0\rangle|10\rangle\}$. We start from the dark initial state $|0\rangle_{ph}(|01\rangle - |10\rangle)_{at}$, which is a coherent antisymmetric superposition of the single-excitation atomic states. Thermal dephasing acts as a ``which-atom'' measurement: it primarily suppresses the off-diagonal coherence between $|01\rangle_{at}$ and $|10\rangle_{at}$, while not directly transferring population between the energy eigenstates.
  
\begin{figure}[H]
   	\includegraphics[width=0.5\textwidth]{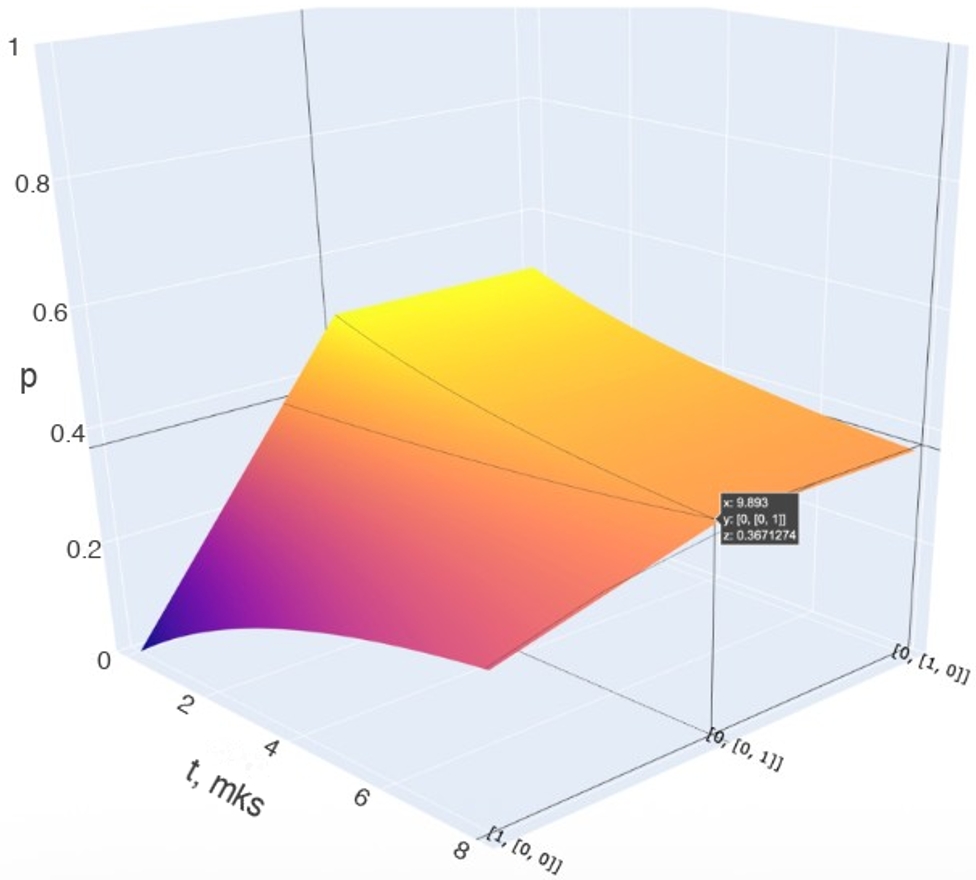}
    \caption{(Online color.) {\it Population dynamics in the truncation $\{|1\rangle|00\rangle$, $|0\rangle|01\rangle$, $|0\rangle|10\rangle\}$ in a shorter time window.}} 
    \label{fig:dephasing_notes_plots_1}
  \end{figure}

\begin{figure}[H]
   	\includegraphics[width=0.5\textwidth]{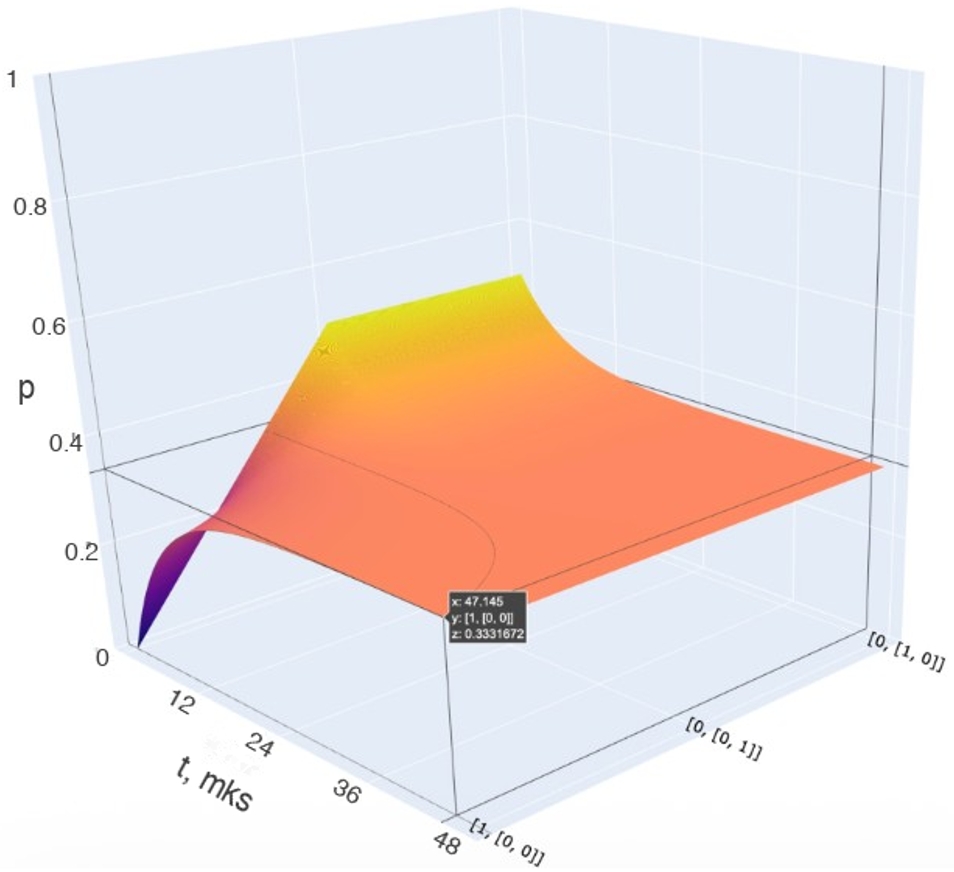}
    \caption{(Online color.) {\it Population dynamics in the truncation $\{|1\rangle|00\rangle$, $|0\rangle|01\rangle$, $|0\rangle|10\rangle\}$ in a longer time window.}} 
    \label{fig:dephasing_notes_plots_2}
  \end{figure}

\begin{figure}[H]
   	\includegraphics[width=0.7\textwidth]{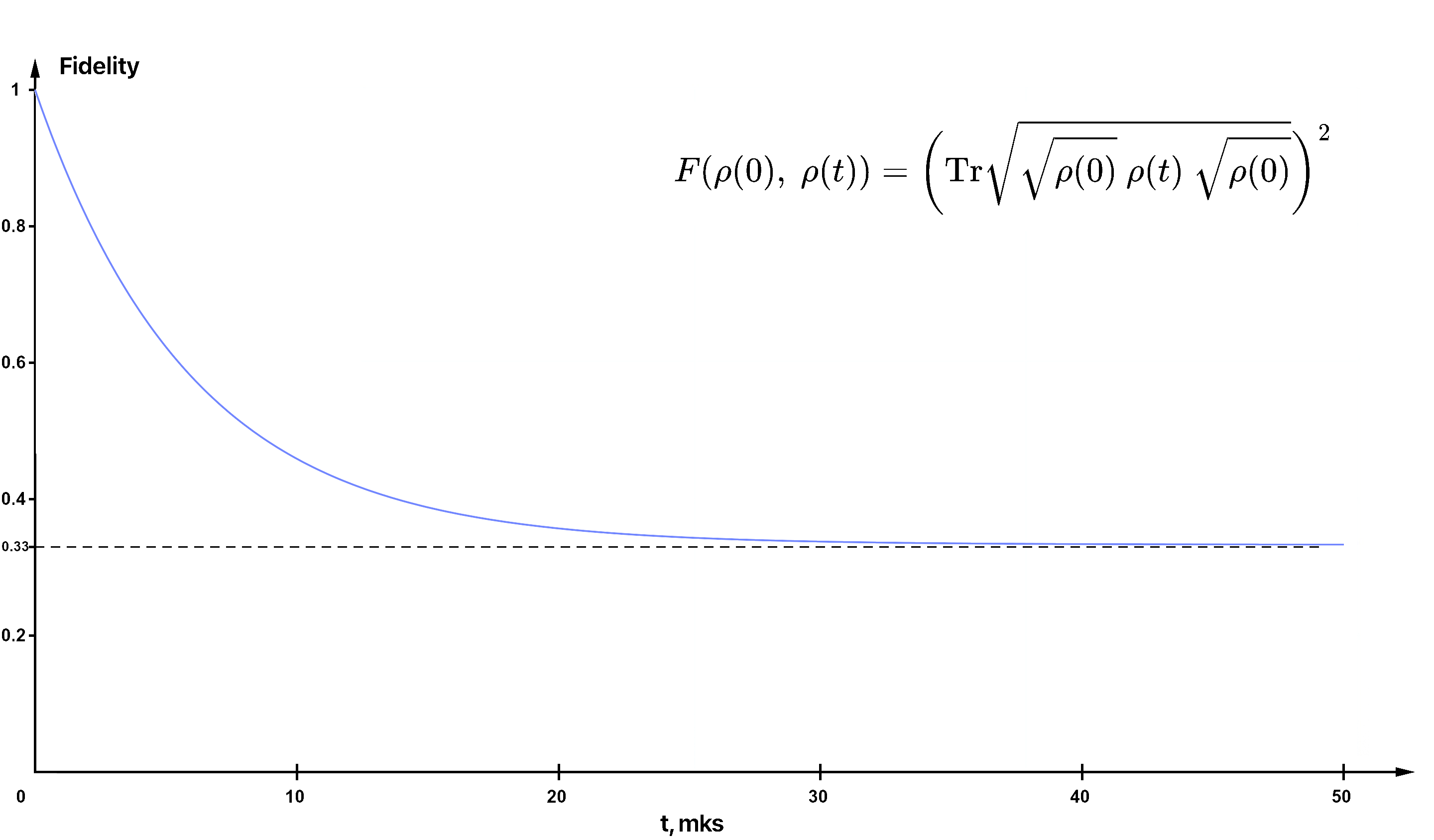}
    \caption{(Online color.) {\it Fidelity $F(\hat{\rho},\hat{\sigma}_D)$, saturating near $\frac{1}{3}$.}}
    \label{fig:dephasing_notes_plots_3}
  \end{figure}

  The two population surfaces in Figures \ref{fig:dephasing_notes_plots_1} and \ref{fig:dephasing_notes_plots_2} visualize the time evolution within the basis $\{|1\rangle|00\rangle$, $|0\rangle|01\rangle$, $|0\rangle|10\rangle\}$. In Figure \ref{fig:dephasing_notes_plots_1}, which shows a shorter time window, the initially pure dark state begins to lose phase coherence. The population then gradually redistributes within the reduced manifold. In Figure \ref{fig:dephasing_notes_plots_2}, which shows a longer time window, the populations approach a quasi-stationary regime close to equal sharing, i.e., $p \approx (1/3, 1/3, 1/3)$. This behavior is conceptually important: in this reduced model, the dephasing noise does not drive the system to a state orthogonal to the dark state. Instead, it washes out coherence and leads to a mixed state that still retains a finite projection onto $|\Psi_D\rangle$. This behavior is seen most clearly in Figure \ref{fig:dephasing_notes_plots_3}. The fidelity $F(\hat{\rho},\hat{\sigma}_D)$ decreases from 1 at $t=0$ but then approaches an asymptote near $0.33\approx 1/3$. In the three-state basis, this asymptote has a simple interpretation: when the reduced manifold approaches an approximately maximally mixed state, $\hat{\rho} \approx I_3/3$, the overlap with any normalized pure state supported on the same manifold is $F(\hat{\rho},\hat{\sigma}_D) = \mathrm{Tr}(\hat{\rho},\hat{\sigma}_D) \approx 1/3$. Therefore, even though dephasing destroys the phase coherence that makes the state perfectly dark, it cannot completely eliminate the dark component in this setting. A finite fraction of the state remains ``dark'' in the sense of having a nonzero projection onto $|\Psi_D\rangle$.
  
  \section{Concluding Discussion and Future Work}
  \label{sec:Conclusion}

	In this work, we have systematically investigated the influence of thermal dephasing---an effective description of continuous QND monitoring---on the stability of dark states in a two-atom cavity QED system. By numerically solving the Lindblad master equation for the TC model, we have mapped out the parameter regimes in which measurement-induced dephasing leads to either accelerated or suppressed dark-state dynamics.

	Our main findings can be summarized as follows. First, we have demonstrated that dark states exhibit a clear crossover from anti-Zeno to Zeno behavior as the dephasing intensity increases. At weak dephasing, increasing the measurement strength reduces both the stabilization time and the dark-state retention, characteristic of the anti-Zeno regime. At stronger dephasing, the trend reverses: the stabilization time grows and the dark-state population recovers, signaling the onset of Zeno-like freezing. Second, we have shown that cavity photon exchange plays a crucial role in this transition. The presence of photon replenishment can partially stabilize the dark-state component, mitigating the destructive effects of dephasing alone. Third, we have analyzed the dependence on asymmetric atom--field couplings, identifying a threshold ratio $\kappa_0$ that separates qualitatively different dynamical regimes. For $\kappa < \kappa_0$, the system exhibits Zeno-first behavior; for $\kappa > \kappa_0$, it follows the anti-Zeno to Zeno pattern; and near $\kappa \approx \kappa_0$, a transitional saddle-like structure emerges. The critical value $\kappa_0$ decreases approximately exponentially with increasing coupling strength.

	Importantly, we have found that even in the strong-dephasing regime, the dark state never becomes completely orthogonal to the target state. In the reduced three-state truncation, the fidelity asymptotically approaches $1/3$, corresponding to a maximally mixed state within the single-excitation manifold. This residual overlap confirms the inherent robustness of dark states against phase noise, a property that persists despite the loss o\mbox{f c}oherence.

	To further contextualize our results, we briefly consider two limiting cases. In a two-atom system without a cavity ($g_1=g_2=0$), the dark state does not exist because there is no cavity-mediated coupling to generate destructive interference. Dephasing would then simply destroy independent atomic coherences, and the Zeno/anti-Zeno behavior would qualitatively differ from that observed here. In a single-atom cavity model (Jaynes--Cummings), the dark-state manifold is absent, and population trapping would occur via different mechanisms (e.g., Zeno freezing of the excited state or dissipative pumping). A detailed comparison with these systems is beyond the present scope, but they represent interesting directions for future research to isolate the role of the dark-state manifold and cavity-mediated entanglement.

	From a broader perspective, our results contribute to the understanding of how continuous monitoring affects coherence in multipartite quantum systems. The observed Zeno--anti-Zeno crossover in dark-state dynamics may have practical implications for quantum information processing, where preserving specific entangled states against decoherence is essential. The ability to partially stabilize dark states through controlled dephasing and photon exchange suggests new strategies for quantum memory and control.

	While our results are obtained for a specific two-atom cavity QED setup, the observed non-monotonic behavior and the frequency-independent steady-state population are expected to be generic features of Zeno/anti-Zeno dynamics in systems with a dark-state~manifold.

	Future work could extend this analysis to larger atomic ensembles, explore the role of correlated dephasing, or investigate the connection to measurement-induced phase transitions. Experimental realization in superconducting circuits or cold-atom platforms would provide a direct test of the predicted phenomena.

  \section*{CRediT authorship contribution statement}
    Conceptualization, Y.I.O. and H.-h.M.; methodology, Y.I.O., R.C., J.Y. and A.V.K.; investigation, R.C., J.Y. and A.V.K.; software, R.C., J.Y. and A.V.K.; data curation, R.C., J.Y. and A.V.K.; visualization, R.C., J.Y. and A.V.K.; formal analysis, R.C., J.Y. and A.V.K.; validation, R.C., J.Y. and H.-h.M.; writing---original draft, R.C. and J.Y.; writing---review and editing, H.-h.M.; project administration, H.-h.M. and Y.I.O.; resources, Y.I.O.; supervision, Y.I.O.
  
  \begin{acknowledgments}
  The reported study was funded by the China Scholarship Council, project numbers 202308091509, 202308091210, and 202108090483. The authors acknowledge the center for collective use of ultra-high-performance computing resources at Lomonosov Moscow State University (\url{https://parallel.ru/}) for providing supercomputer resources that contributed to the research results presented in this paper.
  \end{acknowledgments}
  
\FloatBarrier
\bibliography{bibliography}

\end{document}